\documentclass[a4paper,12pt]{article}
\usepackage[utf8]{inputenc}
\usepackage{amsmath}
\usepackage{tikz}
\usepackage{wrapfig}

\DeclareMathOperator{\Res}{Res}

\DeclareMathOperator{\Arg}{Arg}

\title{Integral formulas for Painlev\'e-2 transcendent.}
\author{O.~M.~Kiselev}

\begin{document}
\maketitle

\begin{abstract}
In the work we use integral formulas for calculating the monodromy data for the Painlev\'e-2 equation.  The perturbation theory for the auxiliary linear system  is constructed and formulas for the variation of the monodromy data  are obtained. We also derive a formula for solving the linearized Painlev\'e-2 equation based on the Fourier-type integral of the squared solutions of the auxiliary linear system of equations.
\end{abstract}

\section{Introduction}

We consider the scheme of the isomonodromic deformation method for the Painleve-2 equation in the following form:
\begin{equation}
u''=2u^3+xu.
\label{eqP2}
\end{equation}

The approach is based on the method of isomonodromic deformations developed in \cite{FlaschkaNewell1980}, \cite{ItsNovokshenov1986}, \cite{Kapaev1988Eng} and \cite{ItsKapaevNovokshenovFokasEng}. Here we obtain integral formulas which allow us to use the method of the isomonodromic deformations to study a variations of the Stockes coefficient and to obtain a formula for the linearized Panleve-2 equation:
\begin{equation}
v''=6u^2v+xv.
 \label{equationLinearizedPainleve2}
\end{equation}
The obtained formulas for solution of (\ref{equationLinearizedPainleve2}) use squared of integrals of solutions for auxiliary linear problem. Such  formulas for solutions of linearized equations are used widely for the perturbed (1+1)-dimensional integrable equations by Inverse Scattering Transform Method for the  first corrections of perturbation theory \cite{Kaup1976}, \cite{KarpmanMaslov1977Eng} and for corrections from the continuous spectrum \cite{Kiselev1992}. For (2+1)-dimensional integrable equations the formulas for linearized equations were obtained in \cite{Kiselev1998}, \cite{Kiselev1998FunkEng}. Here we derive analogously formulas for the theory of integrating of linearized Painlev\'e-2 equation.

The method of constructing the perturbation theory based on the results for the auxiliary scattering problem was successful for both (1+1) - dimensional partial differential equations \cite{Kaup1976}, \cite{KarpmanMaslov1977Eng} and for corrections from the continuous spectrum \cite{Kiselev1992}. Integral formulas for constructing corrections of perturbed equations are also used for (2+1) - dimensional integrable equations \cite{Kiselev1998}, \cite{Kiselev1998FunkEng}.

The approach developed here allows us to study and to obtain formulas for the variations of the Stokes constants, which are the parameters of the Painlev\'e transcendent. These formulas and formula for solution of (\ref{equationLinearizedPainleve2}) open a way to study the properties of the linearized equation using the global properties of the Painlev\'e transcendent.

Here is the general structure of the work.
In section \ref{secIntegralFormulasForStockesConstants}, we present the Stokes theory for solutions of the auxiliary system of equations and derive integral formulas for Stokes matrices. In the section \ref{secIntegralFormulasForPainleveTranscendent} the integral formulas for the Painlev\'e-2 transcendent are obtained using the integral representation of the solution of the Riemann-Hilbert problem for the auxiliary system of equations.  In the section \ref{secVariationsOfStokesCoefficients} the formulas for the variation of the Stokes coefficients are derived. The section \ref{secFormulaForLinearizedPainleve2} provides a formula for solving the linearized Painlev\'e-2 equation.

\section{Integral formulas for the Stokes coefficients}
\label{secIntegralFormulasForStockesConstants}

In this section the integral formulas for solving the auxiliary system of equations for the parameter $\lambda$ are given according to the theory from \cite{FlaschkaNewell1980}, \cite{ItsNovokshenov1986}, \cite{ItsKapaevNovokshenovFokasEng}. These integral formulas are used to obtain integral formulas for the Stokes coefficients of the auxiliary system of equations.

Consider an auxiliary system of equations that determines the dependence of the function $ \ Psi$ on the complex variable $\lambda$:
\begin{eqnarray}
\frac{d\Psi}{d\lambda}=A \Psi,\quad 
A=-i(4\lambda^2+x+2u^2)\sigma_3+ 4u\lambda\sigma_1-2u'\sigma_2.
\label{eqAforP2}
\end{eqnarray}
Here the notation for Pauli matrices is accepted:
\begin{equation}
\sigma_1=\left(\begin{array}{cc} 0 & 1 \\ 1 & 0\\ \end{array}\right),\quad
\sigma_2=\left(\begin{array}{cc} 0&-i\\i&0\\ \end{array}\right),\quad
\sigma_3=\left(\begin{array}{cc} 1&0\\0&-1\\ \end{array}\right).
\label{formulasPMatrix}
\end{equation}

In addition to the system of equations (\ref{eqAforP2}) the function $\Psi$ satisfies the system of differential equations for the real variable $x$:
\begin{equation}
\frac{d \Psi}{d x}=U\Psi,\quad U=-i\lambda \sigma_3+u\sigma_1.
\label{eqUforP2}
\end{equation}

The Painlev\'e-2 equation is a condition for the existence of a solution of both systems of equations (\ref{eqAforP2}) and (\ref{eqUforP2}) \cite{Garnier1912}.

The solution of the system of equations (\ref{eqAforP2}) has the singular point at $\lambda=\infty$. The asymptotics of the solution of this system for $\lambda\to\infty$ can be constructed by the WKB \cite{Wasov1965} method. An explicit form of such an asymptotic  was given in \cite{FlaschkaNewell1980}/ However we  need the asymptotic expansion of the third order of $\lambda^{-1}$:
\begin{eqnarray}
\Psi^\infty\sim
\left(I+
\frac{1}{2\lambda}
\begin{pmatrix}
i(u^2x-(u')^2+u^4) & -iu
\\
iu & -i(u^2-(u')^2+u^4)
\end{pmatrix}
\right.
\nonumber
\\
\left.
\frac{1}{8\lambda^2}
\begin{pmatrix}
p_{11} & p_{21}
\\
p_{21} & p_{11}
\end{pmatrix}
+
\frac{1}{48\lambda^3}
\begin{pmatrix}
q_{11} & -q_{21}
\\
q_{21} & q_{11}
\end{pmatrix}
+O(\lambda^{-4})
\right)
\times 
\nonumber
\\
\times
\exp\left(-i\Omega(\lambda)\sigma_3\right),
\label{formForAsymptoticExpansion}
\end{eqnarray}
where $\Omega(\lambda)= \left(4\lambda^3/3+\lambda x\right)$, the coefficients  $p_{11}$, $p_{21}$, $q_{11}$ and $q_{21}$ are derived by computer algebra system  <<Maxima>>:
\begin{eqnarray*}
p_{11}
&=&
-(u^4 x^2+(2 u^6-2 u^2 (u')^2)x+(u')^4-2u^4(u')^2+u^8-u^2),\\
p_{21}
&=&
-2(u^3 x-u(u')^2-u'+u^5),\\
q_{11}
&=&
i u^6 x^3+(-3iu^4\,(u')^2+3iu^8+2iu^2)\,x^2+
\\
& &
(3iu^2\,(u')^4+(-6iu^6-2i)\,(u')^2+3iu^{10}- iu^4)  x-
\\
& &
 i(u)^6+3iu^4\,(u')^4+(3iu^2-3iu^8)\,(u')^2+
\\ 
& &
2iu\,u' + iu^{12}-3iu^{6},
\\
q_{21}
&=&
-3(i u^5\, x^2+(-2  i u^3\,(u')^2-2iu^2\,u' +2iu^7+2iu)x+ 
\\
& &
i u(u)^4+2(u')^3-2iu^5\,(u')^2-2u^4\,u' +iu^9+iu^3).
\end{eqnarray*}

The main term of this asymptotic oscillates on the lines $\Im (4\lambda^3/3+\lambda x)=0$. In the neighborhood of an infinity, such lines have asymptotes -- straight lines $\arg(\lambda)=\pi (k-1)/3$, $k=1,\dots,6$. For each of these six lines in the neighborhood of infinity, one can define a function $
\Psi_k$ by the given asymptotic direction $\arg (\lambda)=\pi (k-1)/3$:
$$
\Psi_k\sim\Psi^\infty, \quad k=1,2,3,4,5,6.
$$
Since each of the $\Psi_k $ matrices is a fundamental solution system for (\ref{eqAforP2}), so they can be expressed in terms of each The main term of this asymptotic oscillates on the curves $\Im (4\lambda^3/3+\lambda x)=0$. In the neighborhood of an infinity, such curves have asymptotics as straight lines $\arg(\lambda)=\pi (k-1)/3$, $k=1,\dots,6$. These linea are called by Stockes rays. For each of these six lines in the neighborhood of an infinity, you can define a function $
\Psi_k$ by the given asymptotic direction $\arg (\lambda)=\pi (k-1)/3$:
$$
\Psi_k\sim\Psi^\infty, \quad k=1,2,3,4,5,6.
$$
Since each of the $\Psi_k $ matrices is a fundamental solution system for (\ref{eqAforP2}), so they can be expressed in terms of each other:
\begin{equation}
\Psi_{k+1}=\Psi_k S_k.
\label{formulaForConnectionOfFundamentalSolutions}
\end{equation}
Here $S_k$ is a matrix consisting of parameters that depend on the solution of the Painlev\'e-2 equation, but do not depend on the parameter $\lambda$. These $S_k $ matrices are called Stokes matrices. The symbols correspond to those used in the book \cite{ItsKapaevNovokshenovFokasEng}.

\begin{figure}[h]
\begin{center}
\begin{tikzpicture}[ultra thick]

\begin{scope}[thin,scale=0.7]
\draw (-2,-3)--(2,3);
\draw (2.1,3.1) node {$\Arg(\lambda)=\pi/3$};
\draw (-2.1,-3.1) node {$\Arg(\lambda)=4\pi/3$};

\draw (2.5,0.1) .. controls (0.5,0.2)  .. (1.8,2.5);
\draw (1.7,2.7) .. controls (0,0.2)  .. (-1.7,2.7);
\draw (-1.7,2.5) .. controls (-0.5,0.2)  .. (-2.7,0.1);
\draw (-2.7,-0.1) .. controls (-0.5,-0.2)  .. (-1.7,-2.4);
\draw (-1.7,-2.7) .. controls (0,-0.2)  .. (1.7,-2.7);
\draw (1.7,-2.3) .. controls (0.5,-0.2)  .. (2.7,-0.1);

\draw (-2,3)--(2,-3);
\draw (-2.1,3.1) node {$\Arg(\lambda)=2\pi/3$};
\draw (2.1,-3.1) node {$\Arg(\lambda)=5\pi/3$};

\draw (-3,0) -- (3,0);
\draw (3.1,0.1) node {$\Re(\lambda)$};
\end{scope}
\end{tikzpicture}
\end{center}
\caption{The Stokes rays at the directions $\pi (k-1)/6$, $k=1,2,3,4,5,6$  and the curves of integrating, which tend to $\infty$ in the following directions: $\infty_{k},\infty_{k+1}$.}
\label{figStokesRaysP2}
\end{figure}
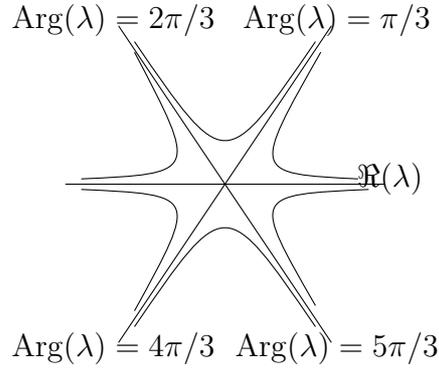

To derive integral formulas for the Stokes matrix, it is convenient to use the substitution:
$$
\Psi_k=\exp\left(-i\Omega(\lambda)\sigma_3\right)\Phi_k.
$$
Using the system of equations (\ref{eqAforP2}) , one can derive a similar system of equations for the matrix$\Phi_k$:

\begin{eqnarray}
\frac{d}{d \lambda}\Phi_k=\left(\exp\left(i\Omega\sigma_3\right)A \exp\left(-i\Omega\sigma_3\right)+i\Omega'\sigma_3\right)\Phi_k.
\label{eqAforPhiForP2}
\end{eqnarray}
For the matrix $\Phi_k$ the following condition is true:
\begin{equation}
\Phi_k\to I,\quad \lambda=R e^{i(k-1)\pi/6},\quad R\to\infty.
\label{asympForPhiK}
\end{equation}

It is not difficult to verify that the solution of the scattering problem (\ref{eqAforPhiForP2}), (\ref{asympForPhiK}) satisfies to a system of integral equations:
\begin{eqnarray}
\Phi_k(\lambda)=I+
\int_{\infty_k}^{\lambda}
\left(\exp\left(i\Omega\sigma_3\right)A \exp\left(-i\Omega\sigma_3\right)+i\Omega'\sigma_3\right)\Phi_k d\mu.
\label{eqIntegrSystem}
\end{eqnarray}
Here the integral is considered as non-proper, where the upper limit is $ \infty_k=R\exp (i\pi (k-1)/3)$, for $R\to\infty$.

Using  (9) one can write the integral over the path from $\infty_k$ to $\infty_{k+1}$:
$$
\Phi_k|_{\lambda\to\infty_{k+1}}=I+
\int_{\infty_k}^{\infty_{k+1}}
\left(\exp\left(i\Omega\sigma_3\right)A \exp\left(-i\Omega\sigma_3\right)+i\Omega'\sigma_3\right)\Phi_k d\mu
\equiv Z_k.
$$
Also as $\lambda\to\infty_{k+1}$:
$$
\Phi_{k+1}|_{\lambda\to\infty_{k+1}}=I,
$$
then
$$
\Phi_{k+1} Z_k=\Phi_k.
$$
The same formula for $\Psi$:
$$
\Psi_{k+1}Z_k=\Psi_k
$$
But a established notation is:
$$
\Psi_{k+1}=\Psi_k S_k.
$$
Let us consider
$$
\Phi_{k+1}|_{\lambda\to\infty_{k}}=I+
\int_{\infty_{k+1}}^{\infty_k}
\left(\exp\left(i\Omega\sigma_3\right)A \exp\left(-i\Omega\sigma_3\right)+i\Omega'\sigma_3\right)\Phi_{k+1} d\mu.
\equiv\tilde{Z}_k
$$
$$
\Phi_{k+1}|_{\infty_k}=\tilde{Z}_k.
$$
But for $\Psi_k$ we have 
$$
\Phi_{k}|_{\infty_k}=I.
$$
Then:
$$
\Phi_{k+1}|_{\infty_k}=\Phi_{k}|_{\infty_k}\tilde{Z}_k,
$$
and $\tilde{Z}_k\equiv S_k$.

According to the formula of the connection of fundamental solutions (\ref{formulaForConnectionOfFundamentalSolutions}) one can obtain:
$$
I+
\int_{\infty_{k+1}}^{\infty_k}
\left(\exp\left(i\Omega\sigma_3\right)A \exp\left(-i\Omega\sigma_3\right)+i\Omega'\sigma_3\right)\Phi_{k+1} d\mu=S_k.
$$
Now $S_k$ can be expressed using $\Psi_k$. 
The integrand in the previous formala can be written as two terms:
\begin{eqnarray*}
\left(\exp\left(i\Omega\sigma_3\right)A \exp\left(-i\Omega\sigma_3\right)+i\Omega'\sigma_3\right)\Phi_{k+1}=
\\
\exp\left(i\Omega\sigma_3\right)A \exp\left(-i\Omega\sigma_3\right)\Phi_{k+1}+i\Omega'\sigma_3\Phi_{k+1}.
\end{eqnarray*}
Now we note:
$$
\exp(-i\Omega\sigma_3)\Phi_{k+1}=\Psi_{k+1}.
$$
Then:
$$
\exp\left(i\Omega\sigma_3\right)A \exp\left(-i\Omega\sigma_3\right)\Phi_{k+1}=\exp\left(i\Omega\sigma_3\right)A\Psi_{k+1},
$$
and
$$
i\Omega'\sigma_3\Phi_{k+1}=
\exp(i\Omega\sigma_3)i\Omega'\sigma_3\exp(-i\Omega\sigma_3)\Phi_{k+1}=\exp(i\Omega\sigma_3)i\Omega'\sigma_3\Psi_{k+1}.
$$
As a result we obtain:
\begin{equation}
S_k=I+
\int_{\infty_{k+1}}^{\infty_k}
\exp\left(i\Omega\sigma_3\right)\left(A+i\Omega'\sigma_3\right)\Psi_{k+1} d\mu.
\label{formulaForSk}
\end{equation}
The integral formula for the matrix $S_k$ is convenient to consider by components. It is important to take into account the asymptotic properties of the matrix $\Psi$ from the formula (\ref{asympForPhiK}).

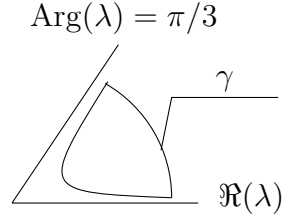
\begin{figure}[h]
\begin{center}
\begin{tikzpicture}[ultra thick]

\begin{scope}[thin,scale=0.7]
\draw (0,0)--(2,3);
\draw (0,0)--(3.5,0);
\draw (2.1,3.5) node {$\Arg(\lambda)=\pi/3$};
\draw (3,0.1) .. controls (0.5,0.2)  .. (1.8,2.3);
\draw (3,0.1) arc [start angle =0,end angle=60,radius=2.5];
\draw (4.5,0.1) node {$\Re(\lambda)$};
\draw (2.8,1)--(3,2)--(5,2);
\draw (4,2.3) node{$\gamma$};
\end{scope}
\end{tikzpicture}
\end{center}
\caption{ The integrals over closed curve  $\gamma$  in the formula(\ref{formulaForSk}) are equal zero due to the Cauchy theorem for analytic functions.}
\label{figCauchyThoerem}
\end{figure}

The integrands in the diagonal elements are analytic functions with repect to  $\lambda$ and the integrands decrease as $\lambda^{-2}$ for $\lambda\to\infty$. Therefore we use the Cauchy theorem for such functions and consider the integrals over an arc of a large circle with radius $R$ as $R\to\infty$ (see. fig. \ref{figCauchyThoerem}).

Let us consider carefully calculations for the matrix  $S_1$ as an example.  The function $\Psi_2$ oscillates near the ray  $Arg(\lambda)=\pi/3$ in the sector  $0<\Arg(\lambda)<2\pi/3$, and the asymptotics of this function  coincides to the asymtotics  (\ref{formForAsymptoticExpansion}) for  $\Psi_{\infty}$. Therefore instead of integrals over the path near the Stockes rays one can use the integrals over an arc of a large circle and the asymptotic behavior of $\Psi_2$.

For others matrix $S_k$ one can use the same calculations in corresponding sectors of the complex plane of $\lambda$. 

Let us consider the integrand on an arc of the the large circle. For this step we calculate 
\begin{eqnarray}
\left(A+i\Omega'\sigma_3\right)=
\left(
\begin{array}{cc}
-i(4\mu^2+x+2u^2) & 4u\mu+2iu'\\
4u\mu -2iu' & i(4\mu^2+x+2u^2)
\end{array}
\right)
+
\nonumber
\\
\left(
\begin{array}{cc}
i4(\mu^2+x) & 0\\
0 & -i(4\mu^2+x)
\end{array}
\right)
=
\left(
\begin{array}{cc}
-2iu^2 & 4u\mu+2iu'\\
4u\mu -2iu' & 2i u^2)
\end{array}
\right).
\label{formulaForAPlusiOmegaPrimeSigma3}
\end{eqnarray}
Now it is conveniet to calculate the multiplication of the mathrix  $A+i\Omega'\sigma_3$ and the left matrix multiplayer withouth the exponent in the asymptotics of $\Psi_\infty$ using  
(\ref{formForAsymptoticExpansion}).

The integrand of $(S_1)_{11}$ looks as follows:
\begin{eqnarray*}
f_{11}
&=& -2iu^2
\bigg(1+\frac{1}{2\mu}\big( i u^2 x- i (u')^2+ iu^4\big)
+
\\
& &
4u\mu
\bigg(
\frac{iu}{2\mu}
-\frac{1}{4\mu^2}(u^3 x-u(u')^2-u'+u^5)
\bigg)+
2iu'
\bigg(
\frac{iu}{2\mu}
\bigg)
\\
&=&
(-2iu^2+2iu^2)+
\\
& &
\frac{1}{\mu}
\big(
u^4x-u^2 (u')^2+u^6-u^4x+u^2(u')^2+uu'-u^6 -uu'
\big)+
\\
& &
O(\mu^{-2})=
O(\mu^{-2}).
\end{eqnarray*}

The integrang in the formula for the coefficient  $(S_1)_{22}$ can be derived by the same way. 

In the integrand for  $(S_1)_{12}$ and $(S_1)_{21}$ the form of the exponent is important. Such simple calculation we do not show here.
As a result we get:
$$
(S_k)_{11}=1+\lim_{R\to\infty}\int_{R\exp(i\pi (k+1)/3)}^{R\exp(i\pi k/3)}
-\frac{i {{u}^{2}} x-i {{w}^{2}}+ i {{u}^{4}}}{2 {{\mu }^{2}}}+\mathcal{O}(R^{-3}) d\mu=1.
$$
$$
(S_k)_{22}=1+\lim_{R\to\infty}\int_{R\exp(i\pi (k+1)/3)}^{R\exp(i\pi k/3)}
\frac{ {{u}^{2}} x- i {{w}^{2}}+ i {{u}^{4}}}{2 {{\mu }^{2}}}+\mathcal{O}(R^{-3}) d\mu=1.
$$

The elements of the matrix $S_k$ lying on the inverse diagonal have exponents in the integrand for large values of $ \ lambda$:
\begin{eqnarray}
(S_k)_{12}=\lim_{R\to\infty}\int_{R\exp(i\pi (k+1)/3)}^{R\exp(i\pi k/3)}
(4 i u \mu +\left( 2 u^3 x-2 uw^2-2 w+2 u^5\right)+
\nonumber
\\
\mathcal{O}(1/R))\exp(2i(4\mu^3/3+x\mu))  d\mu
\label{formulaSk12Asymp}
\\
(S_k)_{21}=\lim_{R\to\infty}\int_{R\exp(i\pi (k+1)/3)}^{R\exp(i\pi k/3)}
(-4 i u \mu +\left( 2 {{u}^{3}} x-2 u\, {{w}^{2}}-2 w+2 {{u}^{5}}\right)+
\nonumber
\\
\mathcal{O}(1/R))
\exp(-2i(4\mu^3/3+x\mu))  d\mu
\label{formulaSk21Asymp}
\end{eqnarray}

The values of the integrals in the formulas (\ref{formulaSk12Asymp}) and (\ref{formulaSk21Asymp}) depend on the sign $\Re(i\mu^3)$ on the integration path. Therefore, it is convenient to make calculations for different values of $k$.

If $k=1,3,5$, then on the arc $\pi (k-1)/3<\arg(\mu)<\pi k/3$ we get $\Re(i\mu^3)=-R\sin(3\arg(\mu))<0$. In this case, it can be shown that
$$
(S_k)_{12}=0.
$$
Similarly, for $k=2,4,6$  on the arc $\pi (k-1)/3<\arg(\mu)<\pi k/3$ we get $\Re(-i\mu^3)=R\sin(3\arg(\mu))<0$, that is:
$$
(S_k)_{21}=0.
$$
As a result, we get:
$$
S_k=
\begin{pmatrix}
1 & 0\\ s_k &1
\end{pmatrix},\quad k=1,3,5;
$$

$$
S_k=
\begin{pmatrix}
1 & s_k\\ 0 &1
\end{pmatrix},\quad k=2,4,6.
$$

In the terms of \cite{FlaschkaNewell1980} we rewrite:
$s_1=a,\ s_2=b,\ s_3=c,\ s_4=d,\ s_5=e,\ s_6=f$ and $s_1=s_4,\ s_2=s_5,\ s_3=s_6$.

Formulas for $s_k$ one can obtain using the maultiplication of the matrix in the integrands in (\ref{formulaForSk}) and formula  (\ref{formulaForAPlusiOmegaPrimeSigma3}):
\begin{eqnarray*}
\left(
\begin{array}{cc}
e^{i\Omega} & 0\\
0 & e^{-i\Omega}
\end{array}
\right)
\left(
\begin{array}{cc}
-2iu^2 & 4u\mu+2iu'\\
4u\mu -2iu' & 2i u^2)
\end{array}
\right)
\left(
\begin{array}{cc}
(\Psi_k)_{11} & (\Psi_k)_{12}\\
(\Psi_k)_{21}& (\Psi_k)_{22}
\end{array}
\right)
=
\\
\left(
\begin{array}{cc}
e^{i\Omega(\mu)} & 0\\
0 & e^{-i\Omega(\mu)}
\end{array}
\right)
\times
\\
\left(
\begin{array}{cc}
-2iu^2(\Psi_k)_{11} + (4u\mu+2iu')(\Psi_k)_{21}
&
-2iu^2(\Psi_k)_{12} + (4u\mu+2iu')(\Psi_k)_{22}
\\
(4u\mu -2iu)(\Psi_k)_{11} + 2i u^2(\Psi_k)_{21}
&
(4u\mu -2iu)(\Psi_k)_{12} + 2i u^2(\Psi_k)_{22}
\end{array}
\right).
\end{eqnarray*}
Now one should use $\Omega=(4/3)\mu^3+\mu x$.

The main result of the section \ref{secIntegralFormulasForStockesConstants} -- explicit formulas for monodromy data:
$$
s_k=2 \int_{\infty_{k+1}}^{\infty_k}
\left((2u\mu -iu')(\Psi_k)_{11}+ i u^2 (\Psi_k)_{21}\right)e^{-i( 4\mu^{3}/3+x \mu) }
d\mu,\quad k=1,3,5; 
$$
$$
s_k=2\int_{\infty_{k+1}}^{\infty_k}
\left((2u \mu + iu')(\Psi_k)_{22}- i u^2 (\Psi_k)_{12}\right)e^{i(4\mu^{3}/3+x \mu) }
d\mu,\quad k=2,4,6.
$$

\section{Integral formula for the Painlev\'e transcendent}
\label{secIntegralFormulasForPainleveTranscendent}

The analytical properties of the functions $\Psi_k$ allow us to formulate the problem of conjugation of functions for the analytical continuation of the function $\Psi_k$ into neighboring sectors of the complex plane of the parameter $\lambda$. To obtain integral equations in this case, it is convenient to use the formulas of Sokhotsky \cite{Sokhotskii1873Eng}. Such constructions were made, in particular, in the work \cite{FlaschkaNewell1980}. As a result, we obtained a system of equations for the first and second columns of analytical equations in the complex plane $ \ lambda$:

\begin{eqnarray*}
\Psi^{(1)}e^{i\Omega}=\begin{pmatrix}1\\0\end{pmatrix}-
\Res_{\mu=0}\frac{\Psi^{(1)}e^{i\Omega}}{\mu-\lambda}+
\frac{s_1}{2\pi i}\int_{C_{42}}\frac{\Psi^{(2)}e^{i\Omega}}{\mu-\lambda}d\mu +
\nonumber
\\
\frac{s_2}{2\pi i}\int_{C_{46}}\frac{\Psi^{(2)}e^{i\Omega}}{\mu-\lambda}d\mu +
\frac{s_2 s_3}{2\pi i}\int_{C_{64}}\frac{\Psi^{(1)}e^{i\Omega}}{\mu-\lambda}d\mu,
\\ 
\Psi^{(2)}e^{-i\Omega}=\begin{pmatrix}0\\1\end{pmatrix}-
\Res_{\mu=0}\frac{\Psi^{(2)}e^{-i\Omega}}{\mu-\lambda}+
\frac{s_2}{2\pi i}\int_{C_{53}}\frac{\Psi^{(1)}e^{-i\Omega}}{\mu-\lambda}d\mu +
\\
\frac{s_3}{2\pi i}\int_{C_{51}}\frac{\Psi^{(1)}e^{-i\Omega}}{\mu-\lambda}d\mu +
\frac{s_1 s_2}{2\pi i}\int_{C_{53}}\frac{\Psi^{(2)}e^{-i\Omega}}{\mu-\lambda}d\mu.
\end{eqnarray*}

The solution of the Painlev\'e-2 equation is usually represented using the asymptotics at $\lambda\to\infty$ for the components of the matrix $\Psi$ lying on the inverse diagonal \cite{FlaschkaNewell1980}:
$$
u(x)=\lim_{\lambda\to\infty}\lambda i \Psi_{12}e^{-i\Omega}
$$
or
$$
u(x)=-\lim_{\lambda\to\infty}\lambda i\Psi_{21}e^{i\Omega}.
$$
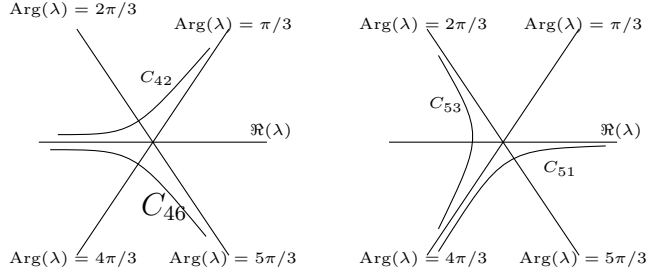
\begin{figure}[h]
\begin{center}
\begin{tikzpicture}[ultra thick,scale=0.5]
\begin{scope}[thin]
\draw (-2,-3)--(2,3);
\draw (2.1,3.1) node {\tiny$\Arg(\lambda)=\pi/3$};
\draw (-2.1,-3.1) node {\tiny$\Arg(\lambda)=4\pi/3$};
\draw (1.5,2.5) .. controls (-0.5,0.2)  .. (-2.5,0.2);
\draw (0.1,1.7) node {\tiny $C_{42}$};
\draw (-2.7,-0.2) .. controls (-0.5,-0.2)  .. (1.4,-2.5);
\draw (0.3,-1.7) node {$C_{46}$};
\draw (-2,3)--(2,-3);
\draw (-2.1,3.5) node {\tiny $\Arg(\lambda)=2\pi/3$};
\draw (2.1,-3.1) node {\tiny $\Arg(\lambda)=5\pi/3$};
\draw (-3,0) -- (3,0);
\draw (3.1,0.3) node {\tiny $\Re(\lambda)$};
\end{scope}
\end{tikzpicture}
\quad
\begin{tikzpicture}[ultra thick,scale=0.5]
\begin{scope}[thin]
\draw (-2,-3)--(2,3);
\draw (2.1,3.1) node {\tiny $\Arg(\lambda)=\pi/3$};
\draw (-2.1,-3.1) node {\tiny $\Arg(\lambda)=4\pi/3$};
\draw (-1.7,2.3) .. controls (-0.5,0.2)  .. (-1.7,-2.3);
\draw (-1.5,1) node {\tiny $C_{53}$};
\draw (-1.7,-2.9) .. controls (0,-0.2)  .. (2.7,-0.1);
\draw (1.5,-0.7) node {\tiny $C_{51}$};
\draw (-2,3)--(2,-3);
\draw (-2.1,3.1) node {\tiny $\Arg(\lambda)=2\pi/3$};
\draw (2.1,-3.1) node {\tiny $\Arg(\lambda)=5\pi/3$};
\draw (-3,0) -- (3,0);
\draw (3.1,0.3) node {\tiny $\Re(\lambda)$};
\end{scope}
\end{tikzpicture}
\end{center}
\caption{Integration paths in the Riemann problem for the matrix function $ \ Psi$ and for calculating the Painlev\'e transcendent by integral formulas.}
\end{figure}
If we use the integral equations, for the matrix $\Psi$, then we can get another expression for the second transcendent of Painlev\'e, through the components of the functions $\Psi_k$.

For this we use the following formula:
$$
\frac{1}{1-q}=1+\frac{q}{1-q}.
$$
If we denote $q=\mu/\lambda$, then:

\begin{eqnarray*}
\lim_{\lambda\to\infty}\lambda \int_{a}^{b}\frac{f(\mu)}{\mu-\lambda}d\mu=\lim_{\lambda\to\infty}\frac{\lambda}{\lambda}
\int_{a}^{b}\frac{f(\mu)}{\frac{\mu}{\lambda}-1}d\mu=
\\
-\int_{a}^{b}f(\mu)d\mu - 
\int_{a}^{b}f(\mu)\frac{\frac{\mu}{\lambda}}{\frac{\mu}{\lambda}-1}d\mu=
\\
-\int_{a}^{b}f(\mu)d\mu - 
\frac{1}{\lambda}\int_{a}^{b}\frac{f(\mu)\mu}{\frac{\mu}{\lambda}-1}d\mu.
\end{eqnarray*}
If
$$
\left|\int_{a}^{b}f(\mu)\mu d\mu\right|<\infty,
$$
then
$$
\lim_{\lambda\to\infty}\lambda \int_{a}^{b}\frac{f(\mu)}{\mu-\lambda}d\mu=-\int_{a}^{b}f(\mu)d\mu.
$$
In our case the interval of the integration is unbounded. The integrals should be considered as inproper. On the Stockes curves the exponents in the integrands are oscillate and therefore the primary terms of the asymptotics as  $\mu\to\infty$ are bounded/ Let us consider:
$$
\lim_{\mu\to\infty_k}(\Psi_k)_{22}\mu\exp(i\Omega(\mu))\sim \mu\exp(2i((4/3)\mu^3+\mu x)).
$$
The oscillated integral with such integrands exists. It yields the same formulas as on the bounded interval.

For one more integral we get:
$$
\lambda\int_0^{\infty_4}(\Psi_4)_{21}\exp(i\Omega)d\mu.
$$
The asymptotic behavior of the integrand:
$$
\lambda\frac{(\Psi_4)_{21}\exp(i\Omega)}{\mu-\lambda}\sim
\frac{\lambda}{\mu-\lambda}\frac{iu}{2\mu}
$$
Then the integral can be represented in the following form:
\begin{eqnarray*}
\lambda\int_0^{\infty_4}\frac{(\Psi_4)_{21}\exp(i\Omega)}{\mu-\lambda}d\mu+\lambda\int_{\infty_6}^0\frac{(\Psi_4)_{21}\exp(i\Omega)}{\mu-\lambda}d\mu=
\\
\lambda\int_{-a}^{\infty_4}
\bigg(\frac{(\Psi_4)_{21}\exp(i\Omega)}{\mu-\lambda}
-\frac{1}{\mu-\lambda}\frac{iu}{2\mu}\bigg)
d\mu+
\\
\lambda\int_{\infty_6}^{a\exp(5i\pi/3)}
\bigg(
\frac{(\Psi_4)_{21}\exp(i\Omega)}{\mu-\lambda}-
-\frac{1}{\mu-\lambda}\frac{iu}{2\mu}
\bigg)
d\mu+
\\
\lambda\int_{-a}^{\infty_4}
\bigg(\frac{1}{\mu-\lambda}\frac{iu}{2\mu}\bigg)
d\mu+
\lambda\int_{\infty_6}^{a\exp(5i\pi/3)}
\bigg(\frac{1}{\mu-\lambda}\frac{iu}{2\mu}
\bigg)
d\mu
+
\\
\end{eqnarray*}
As a result we get:
\begin{eqnarray}
u(x)=-
\frac{s_1}{2\pi}\int^{\infty_4}_{0}(\Psi_4)_{21}e^{i\Omega}d\mu -
\frac{s_1}{2\pi}\int_{\infty_2}^{0}(\Psi_2)_{21}e^{i\Omega}d\mu -
\nonumber
\\
\frac{s_2}{2\pi}\int^{\infty_4}_{0}(\Psi_4)_{21}e^{i\Omega}d\mu -
\frac{s_2}{2\pi}\int_{\infty_6}^{0}(\Psi_6)_{21}e^{i\Omega}d\mu -
\nonumber
\\
\frac{s_2 s_3}{2\pi} \int^{\infty_4}_{0}(\Psi_4)_{21}e^{i\Omega}d\mu -
\frac{s_2 s_3}{2\pi} \int_{\infty_6}^{0}(\Psi_6)_{21}e^{i\Omega}d\mu.
\label{formulaForTrancendentP2ThoughIntegral21}
\end{eqnarray}
Another expression can be obtained by using the $\Psi_{12}$ component:
\begin{eqnarray}
u(x)=
\frac{s_2}{2\pi}\int^{\infty_5}_{0}(\Psi_5)_{12}e^{-i\Omega}d\mu +
\frac{s_2}{2\pi}\int_{\infty_3}^{0}(\Psi_3)_{12}e^{-i\Omega}d\mu +
\nonumber
\\
\frac{s_3}{2\pi}\int^{\infty_5}_{0}(\Psi_5)_{12}e^{-i\Omega}d\mu +
\frac{s_3}{2\pi}\int_{\infty_1}^{0}(\Psi_1)_{12}e^{-i\Omega}d\mu +
\nonumber
\\
\frac{s_1 s_2}{2\pi} \int^{\infty_5}_{0}(\Psi_5)_{12}e^{-i\Omega}d\mu +
\frac{s_1 s_2}{2\pi} \int_{\infty_3}^{0}(\Psi_3)_{12}e^{-i\Omega}d\mu.
\label{formulaForTrancendentP2ThoughIntegral12}
\end{eqnarray}

The formulas (\ref{formulaForTrancendentP2ThoughIntegral21}) and (\ref{formulaForTrancendentP2ThoughIntegral12}) are the main result of the section \ref{secIntegralFormulasForPainleveTranscendent}.

\section{Variation of the Stokes coefficients}
\label{secVariationsOfStokesCoefficients}

Consider the effect of perturbations on the Stokes coefficients associated with the scattering problem (\ref{eqAforP2}). For an infinitesimal perturbation of the coefficients of the system (\ref{eqAforP2}) $u=u+ \delta u$, we obtain a system of equations for the variation of $ \delta\Psi$:
\begin{equation}
\frac{d\delta\Psi}{d\lambda}=A \delta\Psi+\delta A\Psi,\quad 
\delta A=-i(4\lambda^2+x+4u\delta u)\sigma_3+ 4\delta u\lambda\sigma_1-2\delta u'\sigma_2.
\label{eqAforDeltaP2}
\end{equation}
The general solution of the system of equations (\ref{eqAforDeltaP2}) can be represented as:
\begin{eqnarray}
\delta\Psi=\Psi C +\Psi\int \Psi^{-1}\delta A\Psi d\mu, 
\label{intEqAforDeptaP2}
\end{eqnarray}
where $C$ is a matrix composed of arbitrary constants, which are  parameters of the solution of the system (\ref{eqAforDeltaP2}).
Consider the matrix in the integrand of (\ref{intEqAforDeptaP2}):
\begin{eqnarray*}
(\Psi^{-1}\delta A\Psi)_{11}
&=&
\left(\Psi_{2,1}\,\Psi_{2,2}-\Psi_{1,1}\,\Psi_{1,2}\right) 
4\lambda\delta u  
\\
&+&
\left(\Psi_{2,1}\,\Psi_{2,2}+\Psi_{1,1}\,\Psi_{1,2}\right) \, 2i\frac{d}{d x} \delta u 
\\
&-&
\left(\Psi_{1,1}\,\Psi_{2,2}+\Psi_{1,2}\,\Psi_{2,1}\right)4iu\delta u,
\end{eqnarray*}
\begin{eqnarray*}
(\Psi^{-1}\delta A\Psi)_{21}
&=&
\left(\Psi_{1,1}^{2}-\Psi_{2,1}^{2}\right)4\lambda \delta u  -\left(\Psi_{2,1}^{2}+\Psi_{1,1}^{2}\right)2i\left( \frac{d}{d x} \delta u\right) 
\\
&+&
8 i\Psi_{1,1}\, \Psi_{2,1} u v,
\end{eqnarray*}
\begin{eqnarray*}
(\Psi^{-1}\delta A\Psi)_{12}
&=&
\left(\Psi_{2,2}^{2}-\Psi_{1,2}^{2}\right) 4\lambda\delta u  +\left(\Psi_{2,2}^{2}+\Psi_{1,2}^{2}\right)2i\left( \frac{d}{d x} \delta u\right) 
\\
&-&
8 i\Psi_{1,2}\,\Psi_{2,2}u\delta u,
\end{eqnarray*}
\begin{eqnarray*}
(\Psi^{-1}\delta A\Psi)_{22}
&=&
\left(\Psi_{1,1}\,\Psi_{1,2}- \Psi_{2,1}\,\Psi_{2,2}\right)
4\lambda\delta u  
\\
&-&
\left(\Psi_{2,1}\,\Psi_{2,2}+\Psi_{1,1}\,\Psi_{1,2}\right) \, 2i\left( \frac{d}{d x} \delta u\right)
\\
&+&
\left(\Psi_{1,1}\,\Psi_{2,2}+\Psi_{1,2}\,\Psi_{2,1}\right)4iu \delta u.
\end{eqnarray*}

The infinitesimal variation $\delta\Psi$ is used for calculation of variations of the Stokes coefficients. Namely we get:
$$
\delta\Psi_{k+1}=\delta\Psi_k S_k+\Psi_k\delta S_k,
$$
as $\lambda\to\infty_k$ we obtain the condition:
$$
\delta\Psi_k=0,\quad \lambda\to\infty_k.
$$
Then:
\begin{equation}
\delta\Psi_{k+1}\sim\exp(-i\Omega\sigma_3)\delta S_k,\quad \lambda\to\infty_k.
\label{formual1FordeltaPsiKPlus1AsLambdaInftyk}
\end{equation}
In other side one obtains:
$$
\delta\Psi_{k+1}\sim\Psi_{k+1}\int_{\infty_{k+1}}^{\infty_k}\Psi_{k+1}^{-1}\delta A\Psi_{k+1} d\mu, \quad \lambda\to\infty_k.
$$
Let us change
$$
\Psi_{k+1}=\Psi_k S_k\sim
\exp(-i\Omega\sigma_3)S_k, \quad\lambda\to\infty_k,
$$
then:
\begin{equation}
\delta\Psi_{k+1}\sim\exp(-i\Omega\sigma_3)S_k\int_{\infty_{k+1}}^{\infty_k}\Psi_{k+1}^{-1}\delta A\Psi_{k+1} d\mu, \quad \lambda\to\infty_k.
\label{formual2FordeltaPsiKPlus1AsLambdaInftyk}
\end{equation}
Equate  (\ref{formual1FordeltaPsiKPlus1AsLambdaInftyk}) and  (\ref{formual2FordeltaPsiKPlus1AsLambdaInftyk}), multiply  at left side on $\exp(i\Omega\sigma_3)$. It yields:
\begin{equation}
\delta S_k=S_k\int_{\infty_{k+1}}^{\infty_k}\Psi_{k+1}^{-1}\delta A\Psi_{k+1} d\mu.
\label{formula1ForDeltaSk}
\end{equation}
Only one element of the  matrix $S_k$ depends on  $u$ and $u'$. This element is denotes by $s_k$. Then:
$$
\delta S_k=\left(
\begin{array}{cc}
0&0\\
\delta s_k&0
\end{array}
\right),\quad k=1,3,5. 
$$

Denote:
$$
\int_{\infty_{k+1}}^{\infty_k}\Psi_{k+1}^{-1}\delta A\Psi_{k+1} d\mu=
\left(
\begin{array}{cc}
a&b\\
c&d
\end{array}
\right).
$$
Then the formula (\ref{formula1ForDeltaSk}) has the following form:
$$
\left(
\begin{array}{cc}
0&0\\
\delta s_k&0
\end{array}
\right)=
\left(
\begin{array}{cc}
1&0\\
s_k&1
\end{array}
\right)
\left(
\begin{array}{cc}
a&b\\
c&d
\end{array}
\right),
$$
or
$$
\left(
\begin{array}{cc}
0&0\\
\delta s_k&0
\end{array}
\right)=
\left(
\begin{array}{cc}
a&b\\
a s_k +c&b s_k+d
\end{array}
\right).
$$
It yields: $a=0$, $b=0$, $d=0$ and $\delta s_k=c$, where 
$$
c=\int_{\infty_{k+1}}^{\infty_k}
\left(
\left(\Psi_{1,1}^{2}-\Psi_{2,1}^{2}\right)4\lambda \delta u  -\left(\Psi_{2,1}^{2}+\Psi_{1,1}^{2}\right)2i\left( \frac{d}{d x} \delta u\right)+
8 i\Psi_{1,1}\, \Psi_{2,1} u \delta u,\right)
d\mu.
$$
Then the integrals of the diagonal elements along the integration paths marked in Fig. \ref{figStokesRaysP2} gives zeros.
It is convenient to define:
\begin{eqnarray*}
\psi_1^+=\Psi_{11}^2+\Psi_{21}^2,\quad \psi_1^-=\Psi_{11}^2-\Psi_{21}^2,\quad 
\psi_1=\Psi_{11}\Psi_{21};
\\
\psi_2^+=\Psi_{12}^2+\Psi_{22}^2,\quad \psi_1^-=\Psi_{22}^2-\Psi_{12}^2,\quad 
\psi_2=\Psi_{12}\Psi_{22};
\end{eqnarray*}

For reasons that completely repeat the calculations of the Stokes coefficients from the section \ref{secIntegralFormulasForStockesConstants}, we get:
\begin{eqnarray}
\delta s_k
&=&
\int_{\infty_{k+1}}^{\infty_{k}} 4\mu\delta u\, \psi_1^- 
-
2i\left( \frac{d}{d x} \delta u\right)\,\psi_1^+ 
+
8 i u\delta u\, \psi_1d\mu,\, k=1,3,5; 
\label{formDeltaS135}
\\
\delta s_k
&=&
\int_{\infty_{k+1}}^{\infty_{k}}
4\mu\delta u \,\psi_2^- 
+
2i\left( \frac{d}{d x} \delta u\right)\, \psi_2^+ 
-
8 iu\delta u\,\psi_2 d\mu,\, k=2,4,6.
\label{formDeltaS246}
\end{eqnarray}

\section{Formula for solution of the linearized Painlev\'e-2 equation}
\label{secFormulaForLinearizedPainleve2}

Consider the equations for quadratic expressions $
\psi_1^+,\, \psi_1^-,\, \psi_1$. To do this, it is convenient to use the system of equations (\ref{eqAforP2}):\begin{eqnarray}
\frac{d \Psi_{11}}{d\lambda}
&=&
-i(4\lambda^2+x+2u^2)\Psi_{11}
+
(4u\lambda+2iu')\Psi_{21},
\label{eqA1forP2}
\\
\frac{d \Psi_{21}}{d\lambda}
&=&
(4u\lambda-2iu')\Psi_{11}
+
i(4\lambda^2+x+2u^2)\Psi_{21}
\label{eqA2forP2}
\end{eqnarray}
The differential equation for $\psi_1^+$ by $\lambda$ is obtained if the equation (\ref{eqA1forP2}) is multiplied by $2\Psi_{11}$, the equation (\ref{eqA2forP2}) is multiplied by $2\Psi_{21}$ and the resulting equations are added. As a result, we get:
\begin{equation}
\frac{d \psi^+_1}{d\lambda}=-2i(4\lambda^2+x+2u^2)\psi_1^- + 16\lambda u\psi_1.
\label{eqLambdapsi1ForP2}
\end{equation}

The differential equation for $\psi_1^ - $ by $\lambda$ is obtained if the first equation of the system equation (\ref{eqA1forP2}) is multiplied by $2\Psi_{11}$, the second equation of the system (\ref{eqA2forP2}) is multiplied by $2\Psi_{21}$ and the resulting equations are subtracted. As a result, we get:

\begin{equation}
\frac{d\psi^-_1}{d\lambda}=-2i(4\lambda^2+x+2u^2)\psi_1^+ + 8iu'\psi_1.
\label{eqLambdapsi2ForP2}
\end{equation}

The differential equation for $\psi_1$ by $\lambda$ is obtained if the first equation of the system equation (\ref{eqA1forP2}) is multiplied by $\Psi_{21}$, the second equation of the system (\ref{eqA2forP2}) is multiplied by $\Psi_{11}$ and the resulting equations are added. As a result, we get:
\begin{equation}
\frac{d\psi_1}{d\lambda}=4 \lambda u\,\psi_1^+ -2iu'\,\psi^-_1.
\label{eqLambdapsi3ForP2}
\end{equation}
Similar expressions are obtained for the derivatives of the same quadratic expressions with respect to $x$. To obtain differential equations for $x$, it is convenient to use the second auxiliary system of equations (\ref{eqUforP2} for $\Psi$ or component-by-component:\begin{eqnarray}
\frac{d \Psi_{11}}{d x}
&=&
-i\lambda\Psi_{11}
+u\Psi_{21},
\label{eqXForPsi11}
\\
\frac{d \Psi_{21}}{d x}
&=&
u\Psi_{11}
+
i\lambda\Psi_{21}.
\label{eqXForPsi21}
\end{eqnarray}
Similar transformations give a system of differential equations for the variable $x$:
\begin{eqnarray}
\frac{d\psi^+_1}{d x}
&=&
-2i\lambda\psi^-_1 + 4u\psi_1,
\label{eqXpsi1forP2}
\\
\frac{d\psi^-_1}{d x}
&=&
-2i\lambda\psi^+_1,
\label{eqXpsi2forP2}
\\
\frac{d\psi_1}{d x}
&=&
u\psi_1	^+,
\label{eqXpsi3forP2}
\end{eqnarray}
We assume that the variation $ \delta u$ is the solution of the linearized equation:
\begin{equation}
\delta u''=(6u^2+x)\delta u,
\label{eqLinP2}
\end{equation}

Differentiating (\ref{formDeltaS135}) by $x$ by virtue of the equations (\ref{eqXForPsi11}), (\ref{eqXForPsi21}), and the linearized equation (\ref{eqLinP2}) gives (here and below, $s_1$is considered for certainty):
\begin{eqnarray*}
\frac{d\delta s_1}{d x}
&=&
\int^{\infty_{1}}_{\infty_6}
\left((-2i(4\mu^2+x+2u^2)\psi_1^+ +8iu'\psi_1)\delta u-2if\psi_1^+\right)d\mu
\\
&=&
\delta u \int^{\infty_{1}}_{\infty_6}\frac{d\psi^-_1}{d\mu}d\mu.
\end{eqnarray*}
To calculate the integral of the derivative of $\lambda$, let us consider  the representation of $\psi^-_1$ through the squares of the first column of the $\Psi$ -- function.
For  $\lambda\to \infty_{6}$ and   $r=|\lambda|,\, \alpha=\Arg(\lambda)$ we obtain:
\begin{eqnarray*}
\Psi_{11}^2
&\sim &
\exp\left(-2i\left(\frac{4}{3}r^3 e^{3i\alpha}+x r e^{i\alpha}\right)\right),
\\
\Psi_{21}^2
&\sim &
\frac{-u^2 e^{-2i\alpha}}{4r^2}\exp\left(-2i\left(\frac{4}{3}r^3 e^{3i\alpha}+x r e^{i\alpha}\right)\right), \quad \lambda\to \infty_{6},
\end{eqnarray*}

As  $\lambda\to \infty_{1}$ and  $r=|\lambda|,\, \beta=\Arg(\lambda)-\pi/3$ we get:
\begin{eqnarray*}
\Psi_{11}^2
&\sim & 
\exp\left(2i\left(\frac{4}{3}r^3e^{3i\beta}-xre^{i\pi/3}e^{i\beta}\right)\right)(1+O(r^{-1}))+O(1/r)
\\
&+&
s_1^2\exp\left(-2i\left(\frac{4}{3}r^3e^{3i\beta}-xre^{i\pi/3}e^{i\beta}\right)\right)\left(\frac{u^2e^{-2i\pi/3-2i\beta}}{4r^2}+O(r^{-3})\right),
\\
\Psi_{21}^2
&\sim & 
s_1^2\exp\left(-2i\left(\frac{4}{3}r^3e^{3i\beta}-xre^{i\pi/3}e^{i\beta}\right)\right)(1+O(r^{-1}))+O(1/r)
\\
&+&
\exp\left(2i\left(\frac{4}{3}r^3e^{3i\beta}+xre^{i\pi/3}e^{i\beta}\right)\right)\left(\frac{e^{2i\pi/3+2i\beta}}{r^2}+O(r^{-3})\right).
\end{eqnarray*}

The integral of the derivative $\psi^-_1$ by $\lambda$ is written as the sum of the integrals:

\begin{eqnarray*}
\int^{\infty_{1}}_{\infty_6}\frac{d\psi^-_1}{d\mu}d\mu
=
\int_{\mathcal{L}_{11}}\frac{d\psi_{11}^2}{d\mu}d\mu 
+
\int_{\mathcal{L}_{21}}\frac{d\psi_{21}^2}{d\mu}d\mu.
\end{eqnarray*}
For each of the integrals, we deform the contour so that at its ends the functions $\Psi_{11}^2$ and $\Psi_{21}^2$, respectively, vanish. Questions concerning the path of integration for the representation of solution of the auxiliary linear equations connecting the Painlev\'e-2 equation, see \cite{Suleimanov2008Eng}, \cite{JoshiKitaevTreharne2009}.

Let us consider the asyptotics of $\Psi_{11}^2$ near the ray  to $\infty_1$, as $\Arg(\lambda)=\pi/3+\beta$, where $\beta\ll1$ and $r\to\infty$.
The curve, where the real part of the exponent is equal to zero:
$$
\Re\left(2i\left(\frac{4}{3}\lambda^3+x\lambda\right)\right)=0
$$
as $\lambda=re^{i\pi/3}e^{i\beta}$ has a form:
$$
\frac{8}{3} r^2 \sin(3\beta)-x \sin(\beta)-\sqrt{3} x \cos(\beta)=0
$$
As $r\to\infty$ this curve has an asymptotics:
$$
\beta\sim\frac{x\sqrt{3}}{8r^2}+O(r^{-4})
$$

The asymptotics of $\Psi_{11}$ is following: 
$$
\Psi_{11}^2\sim \exp\left(-\frac{8}{3}r^3\beta+rx\beta +\sqrt{3}xr\right)O(1)+
\frac{\exp(\frac{8}{3}r^3\beta-xr\beta-\sqrt{3}xr)}{r^2}O(1).
$$

First term of this formula decreases as  $\beta>x\sqrt{3}/(8r^2)$, second term decreases as:
$$
\left(\frac{8}{3}r^3\beta-xr\beta-\sqrt{3}xr\right)-2\ln(r)<0,
$$
or as $r\to\infty$:
$$
\beta<\frac{\sqrt{3}}{8r^2}x+\frac{1}{4r^3}\ln(r).
$$
Then $\Psi_{11}^2\to0$ as $r\to\infty$ and:
$$
\frac{\pi}{3}+\frac{\sqrt{3}}{8r^2}x<\Arg(\lambda)<\frac{\pi}{3}+\frac{\sqrt{3}}{8r^2}x+\frac{1}{4r^3}\ln(r).
$$  

By the same way one can get the asymptotics of $\Psi_{21}^2$ near the ray  $(0,\infty_1)$, as $\Arg(\lambda)=\pi/3+\beta$, where $\beta\ll1$ and $r\to\infty$
$$
\Psi_{21}^2\sim \frac{\exp\left(-8r^3\beta+xr\beta+\sqrt{3}xr\right)}{r^2}O(1)+
\exp\left(8r^3\beta-xr\beta-\sqrt{3}xr\right)O(1).
$$
For this function the exponents change the sign with respect to the another asymptotics of  $\Psi_{11}^2$. Therefore the same considerations give the condition:  $\Psi_{21}^2\to0$ as $r\to\infty$ and:
$$
\pi/3-\frac{x\sqrt{3}}{8r^2}>\Arg(\lambda) >\pi/3-\frac{x\sqrt{3}}{8r^2}-\frac{1}{4r^3}\ln(r).
$$

\begin{figure}
\begin{tikzpicture}
\begin{scope}[scale=0.8]

\draw[ultra thick] (-3,0) -- (3,0);
\draw (-0.2,0.3) node {$0$};
\draw (4,0.6) node {\small $\Arg(\lambda)=0$};
\draw [<-] (2.7,0.05)--(2.5,0.3)--(5,0.3);
\draw [thick](0,0)--(3,-0.5);
\draw (4.5,-1) node {\small $-\Delta<\Arg(\lambda)<0$};
\draw [<-](1.5,-0.3)--(2.5,-1.3)--(6.5,-1.3);

\draw [ultra thick] (-1,-1.5)--(2,3);
\draw (4.3,3) node {$\Arg(\lambda)=\pi/3$};
\draw[<-] (2,2.8)--(2.5,2.5)--(6,2.5);

\draw (0.8,2.1).. controls (1.25,2.4) .. (1.7,3);
\draw [->](-5,2.8)--(1,2.8)--(1.25,2.5);
\draw (-2,3.2) node {\small $\Arg(\lambda)=\pi/3+\frac{x\sqrt{3}}{8r^2}+\frac{1}{4r^3}\ln(r)$};

\draw (0.5,1.5).. controls (1.5,2.4) .. (2.05,3.2);
\draw [->](-5,1.8)--(0,1.8)--(0.5,1.5);
\draw (-2,2.2) node {\small $\Arg(\lambda)=\pi/3+\frac{x\sqrt{3}}{8r^2}$};

\draw [ultra thick](3,-0.2) .. controls (0.5,0.2)  .. (1.8,3);
\draw (3,1.6) node {\small $\mathcal{L}_{11}$};
\draw[<-] (1,0.5)--(2.5,1.3)--(3.5,1.3);
\end{scope}
\end{tikzpicture}
\caption{ The ends of the path of integration  $\mathcal{L}_{11}$ lie into the sectors $\pi/3+\frac{x\sqrt{3}}{8r^2}<\Arg(\lambda)<\pi/3+\frac{x\sqrt{3}}{8r^2}+\frac{1}{4r^3}\ln(r)$ and $-\Delta<\Arg(\lambda)<0$, for $\forall\Delta>0$.}
\label{figPathOfIntegrating11}
\end{figure}
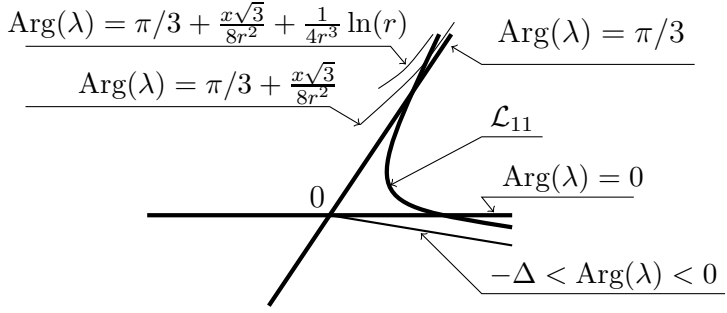

The integral in the formula for the derivation of  $\Psi_{11}^2$ the path of the integration $\mathcal{L}_{11}$ is shown on the figure \ref{figPathOfIntegrating11}. The path begins at the point  $\mathcal{L}_{11}^-$ into the sector $-\Delta<Arg(\lambda)<0$, where $\Delta>0$ and finishes at the point  $\pi/3+\frac{x\sqrt{3}}{8r^2}+\frac{1}{4r^3}\ln(r)>Arg(\mathcal{L}_{11}^+)>\pi/3+\frac{x\sqrt{3}}{8r^2}$, where $r=|\lambda|$.

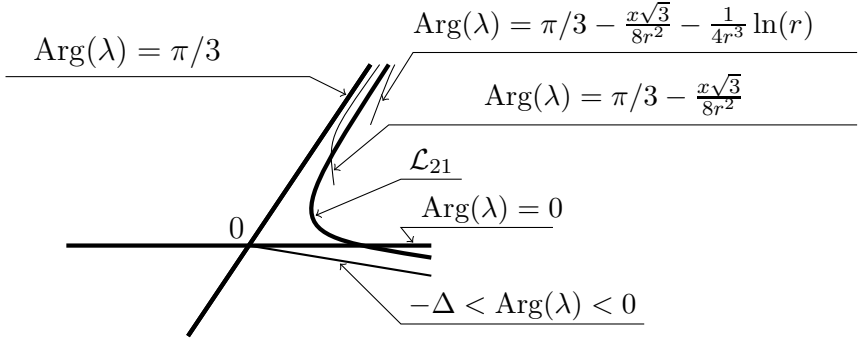
\begin{figure}
\begin{tikzpicture}
\begin{scope}[scale=0.8]

\draw[ultra thick] (-3,0) -- (3,0);
\draw (-0.2,0.3) node {$0$};
\draw (4,0.6) node {\small $\Arg(\lambda)=0$};
\draw [<-] (2.7,0.05)--(2.5,0.3)--(5,0.3);
\draw [thick](0,0)--(3,-0.5);
\draw (4.5,-1) node {\small $-\Delta<\Arg(\lambda)<0$};
\draw [<-](1.5,-0.3)--(2.5,-1.3)--(6.5,-1.3);

\draw [ultra thick] (-1,-1.5)--(2,3);
\draw (6,3.7) node {\small $\Arg(\lambda)=\pi/3-
\frac{x\sqrt{3}}{8r^2}-\frac{1}{4r^3}\ln(r)$};
\draw[<-] (2.2,2.4)--(2.7,3.2)--(10,3.2);
\draw (2,2).. controls(2.3,2.8).. (2.4,3);

\draw [ultra thick] (-1,-1.5)--(2,3);
\draw (6,2.5) node {\small $\Arg(\lambda)=\pi/3-\frac{x\sqrt{3}}{8r^2}$};
\draw[<-] (1.4,1.25)--(2.3,2)--(10,2);
\draw (1.4,1).. controls(1.3,1.8).. (2.15,3);

\draw [->](-4,2.8)--(1,2.8)--(1.6,2.5);
\draw (-2,3.2) node {$\Arg(\lambda)=\pi/3$};

\draw [ultra thick](3,-0.2) .. controls (0.5,0.2)  .. (2.3,3);
\draw (3,1.4) node {\small $\mathcal{L}_{21}$};
\draw[<-] (1.1,0.5)--(2.5,1.1)--(3.5,1.1);
\end{scope}
\end{tikzpicture}
\caption{ The ends of the path of integration $\mathcal{L}_{21}$ lie into the sectors $\pi/3-\frac{x\sqrt{3}}{8r^2}-\frac{1}{4r^3}\ln(r)<\Arg(\lambda)<\pi/3-\frac{x\sqrt{3}}{8r^2}$ and $-\Delta<\Arg(\lambda)<0$ for $\forall\Delta>0$.}
\label{figPathOfIntegrating21}
\end{figure}

The path $\mathcal{L}_{21}$ is obtained using the same considerations. This path is shown in the figure \ref{figPathOfIntegrating21}.

Then the formula is valid:
\begin{equation}
\frac{d\delta s_1}{d x}=0.
\label{eqModS}
\end{equation}

The formulas for the squares of $\Psi$ allow us to represent the solution of the linearized Painleve-2 equation in terms of quadratic expressions from $\Psi$. Indeed, we differentiate by $x$ the equation (\ref{eqXpsi1forP2}) by virtue of the equations (\ref{eqXpsi2forP2}) and (\ref{eqXpsi3forP2}):\begin{eqnarray*}
\frac{d^2\psi^+_1}{d x^2}=4(-\lambda^2+u^2)\psi^+_1+4u'\psi_1.
\end{eqnarray*} 
In this equation, the last term on the right is replaced by the equation (\ref{eqLambdapsi2ForP2}):
$$
4u'\psi_1=(4\lambda^2+x+2u^2)\psi^+_1 - i\frac{1}{2}\frac{d\psi_1^-}{d\lambda}
$$
As a result, we get:
$$
\frac{d^2\psi^+_1}{d x^2}=(x+6u^2)\psi^+_1
- i\frac{1}{2}\frac{d\psi_1^-}{d\lambda}.
$$
The same calculations for $\psi^-_1$ give:
$$
\frac{d^2\psi^-_1}{d x^2}=4(-\lambda^2)\psi^-_1-
8i\lambda u\psi_1.
$$

Replacement:
$$
-8i\lambda\psi_1=-i\frac{1}{2}\frac{d\psi_1^+}{d\lambda}
+(4\lambda^2+x+2u^2)\psi_1^-
$$
yields:
$$
\frac{d^2\psi^-_1}{d x^2}=(x+2u^2)\psi^-_1
-i\frac{1}{2}\frac{d\psi_1^+}{d\lambda}.
$$
These formulas are useful for deriving an integral representation of the solution of the linearized Painleve-2 equation.

Consider the integral:
$$
v(x)=\int_{\infty_{1}}^{\infty_6}\psi^+_1(\lambda,x)d\lambda.
$$
The second derivative of this integral is:
\begin{eqnarray*}
\frac{d^2}{dx^2}\int_{\infty_{6}}^{\infty_1}\psi^+_1(\lambda,x)d\lambda
&=&
(6u^2+x)\int_{\infty_{6}}^{\infty_1}\psi^+_1(\lambda,x)d\lambda-
\\
& &
i\frac{1}{2}\int_{\mathcal{L}_{11}}\frac{d}{d\lambda} \psi_{11}^2d\lambda +
i\frac{1}{2}\int_{\mathcal{L}_{21}}\frac{d}{d\lambda} \Psi_{21}^2d\lambda.
\end{eqnarray*}
Then, for the same reasons as in the derivation of the formula (\ref{eqModS}), we obtain that the solution of the linearized Painlev\'e-2
$$
v''=(6u^2+x)v
$$
can be represented as:
\begin{equation}
v(x)=\int_{\infty_6}^{\infty_1}\psi^+_1(\lambda,x)d\lambda.
\label{formSolLinP2}
\end{equation}

The same way yields us the formula 
$$
y=\int_{\infty_6}^{\infty_1}\psi^-_1(\lambda,x)d\lambda,
$$
which is a solution for:
$$
y''=(x+2u^2)y.
$$

\section{Conclusion}
In this work we obtain the integral formulas for the monodromy data for auxiliary linear equations connection with the Painlev\'e-2 equation. The formulas allow us to derive the perturbation theory for the auxiliary linear system  (\ref{eqAforP2}) and to obtain the formulas for the infinitesimal variations of the monodromy data. Also we derive the integral formula for the solution of the linearized Painlev\'e-2 equation. This formula uses the squares of the solutions of  the auxiliary system of equations (\ref{eqAforP2}).

\section{Acknowledgment}
I thank Bulat Irekovich Suleimanov for his attention and his help during my work on this subject and a lot of comments of the manuscript.

\end{document}